\pdfoutput=1
\documentclass[preprint,10pt]{sigplanconf}
\usepackage{amsmath}
\usepackage{xspace}
\usepackage{filecontents}
\usepackage[dvipsnames,table,xcdraw]{xcolor}
\usepackage{algorithm2e}
\usepackage{graphicx}
\usepackage{caption}
\usepackage{subcaption}
\usepackage{longtable}
\usepackage{ctable}
\usepackage{url}
\usepackage{amsfonts}
\usepackage{pbox}
\usepackage{comment}

\usepackage{listings}          
\lstdefinelanguage{Julia}%
  {morekeywords={abstract,break,case,catch,const,continue,do,else,elseif,%
    end,export,false,for,function,immutable,import,importall,if,in,%
        macro,module,otherwise,quote,return,switch,true,try,type,typealias,%
        using,while,Int,Int64,Float64,Matrix,Vector},%
  ndkeywords={@acc,hpat},%
  ndkeywordstyle=\color{red},%
  otherkeywords={*,\.*,:,-,.,.==,^,=,<,>},
  sensitive=true,%
  alsoother={$},%
  morecomment=[l]\#,%
  morecomment=[n]{\#=}{=\#},%
  morestring=[s]{"}{"},%
  morestring=[m]{'}{'},%
}[keywords,comments,strings]%

\makeatletter
\newcommand{\labitem}[2]{%
\def\@itemlabel{\textbf{#1}}
\item
\def\@currentlabel{#1}\label{#2}}
\makeatother

\newcommand{\Comment}[1]{}
\newcommand{\Space}[1]{} 

\newcommand{\Sys}{{HiFrames}\xspace}
\newcommand{\mpic}{{MPI/C++}\xspace}
\newcommand{\cpp}{{C++}\xspace}
\newcommand{\charm}{{Charm++}\xspace}

\newcommand{\df}{data frame\xspace}
\newcommand{\dfs}{data frames\xspace}
\newcommand{\Juliadf}{{Julia's DataFrames.jl}\xspace}
\newcommand{\hpat}{HPAT\xspace}
\newcommand{\pa}{ParallelAccelerator\xspace}
\newcommand{\sparksql}{Spark SQL\xspace}
\newcommand{\spark}{Spark\xspace}
\newcommand{\sql}{SQL\xspace}
\newcommand{\api}{API\xspace}

\newcommand{\Python}{Python\xspace}
\newcommand{\Julia}{Julia\xspace}
\newcommand{\R}{R\xspace}
\newcommand{\CodeIn}[1]{{\small \texttt{\text{#1}}}}
\newcommand{\CodeInsmall}[1]{{\footnotesize \texttt{\text{#1}}}}
\newcommand{\join}{\CodeIn{join}\xspace}

\newcommand{\map}{\textit{map}\xspace}
\newcommand{\reduce}{\textit{reduce}\xspace}
\newcommand{\cartesianmap}{\textit{Cartesian map}\xspace}
\newcommand{\stencil}{\textit{stencil}\xspace}
\newcommand{\parfor}{\textit{parfor}\xspace}

\newcommand{\dfpass}{\CodeIn{DataFrame-Pass}\xspace}

\definecolor{comment-red}{rgb}{0.8,0,0}

\newcommand{\INTELR}{Intel\textsuperscript{\textregistered}}
\newcommand{\XEONR}{Xeon\textsuperscript{\textregistered}}

\usepackage{lstlinebgrd} 

\begin{document}

\setlength{\pdfpageheight}{\paperheight}
\setlength{\pdfpagewidth}{\paperwidth}

\title{\Sys: High Performance Data Frames in a Scripting Language}


\authorinfo{Ehsan Totoni}
{Intel Labs, USA}
{ehsan.totoni@intel.com}
\authorinfo{Wajih Ul Hassan}
{University of Illinois at Urbana-Champaign, USA}
{whassan3@illinois.edu}
\authorinfo{Todd A. Anderson}
{Intel Labs, USA}
{todd.a.anderson@intel.com}
\authorinfo{Tatiana Shpeisman}
{Intel Labs, USA}
{tatiana.shpeisman@intel.com}

\maketitle

\begin{abstract}
Data frames in scripting languages are essential abstractions for processing structured data.
However, existing data frame solutions are either not distributed (e.g., Pandas in Python) and therefore have limited scalability,
or they are not tightly integrated with array computations (e.g., \sparksql).
This paper proposes a novel compiler-based approach where 
we integrate data frames into the High Performance Analytics Toolkit (HPAT) to build \Sys.
It provides expressive and flexible data frame APIs which
are tightly integrated with array operations.
\Sys then automatically parallelizes and compiles relational operations
along with other array computations in end-to-end data analytics programs,
 and generates efficient \mpic code.
We demonstrate that \Sys is significantly faster than alternatives such as
\sparksql on clusters, without
forcing the programmer to switch to embedded SQL for part of the program.
\Sys is 3.6x to 70x faster than \sparksql
 for basic relational operations,
and can be up to 20,000x faster for advanced analytics
operations, such as weighted moving averages (WMA), that the map-reduce paradigm cannot handle effectively.
\Sys is also 5x faster than \sparksql for TPCx-BB Q26 on 64 nodes of Cori supercomputer.

\end{abstract}

\section{Introduction}\label{sec:intro}
The rise of data science has led to the emergence of a wide variety
of data analytics frameworks that support relational operations within a general purpose
programming language. 
These frameworks can be roughly split into two categories - data frame packages and big data frameworks.
Data frame packages in scripting languages,  
such as \Python Pandas~\cite{mckinneypandas}, \R data frames~\cite{rdataframe}, and \Julia DataFrames~\cite{juliadataframe}
allow quick prototyping of data analytics algorithms. 
Because data frames are typically implemented as collections of their column arrays, they support
standard array operations as well as relational query APIs. 
Scripting data frames have varying levels of
integration with the language but all have the same performance limitations
as the underlying scripting system. Data frame packages run sequentially 
and are limited in the amount of data they can process to what can fit in the memory of a single node.

An alternative approach is based on the map-reduce paradigm~\cite{zaharia2010,dean2008,hadoopweb,impala2015,tezweb,flinkweb}
and distributed execution engines. \sparksql~\cite{armbrust2015} and DryadLINQ~\cite{dryadlinq} provide examples
of such systems.
Big data frameworks allow fault-tolerant processing of large amounts of data on multiple nodes
of a distributed system. But, they suffer from several limitations.
First, they do not provide as seamless and efficient integration between relational
and procedural operations as scripting data frame packages. Because relational operations are implemented
by a separate sub-system via lazy evaluation, they cannot be efficiently integrated with arbitrary
non-relational processing. Second, systems that rely on map-reduce paradigm cannot
efficiently implement distributed computational patterns that go beyond map-reduce, such 
as scan or stencil. As a result, they can be prohibitively slow for advanced analytical
operations, such as computing cumulative sums or moving averages. Finally, distributed
execution via master-slave library approach is known to carry large overheads, as master node
acts as a sequential bottleneck~\cite{HPAT}. 

In this paper, we introduce \Sys in an effort to improve the programmability and performance of relational processing in analytics programs.
\Sys provides a set of scripting
language extensions for structured data processing and a corresponding
framework that can automatically parallelize and compile data analytics programs for distributed execution.
Similar to data frame packages, \Sys provides a data frame interface, supports smooth integration of relational
and analytical processing and requires minimal typing. Similar to big data frameworks, \Sys allows for distributed execution across large data sets. 
In addition, \Sys optimizes across relational and non-relational data-parallel operations, supports flexible communication patterns and leverages
the power of \mpic to provide efficient and scalable distributed execution on small or large clusters. 

\Sys uses a novel compilation approach to generating efficient code for data frame operations. Traditionally, data frames are implemented
as complex objects, such as, for example, arrays of arrays. This representation is necessary to support complex relational operations (join, aggregate)
but limits optimization opportunities with array-based code. In \Sys, each data frame column represented as a separate array variable and all 
data frame operations are expanded to work on individual arrays. This allows \Sys to generate highly efficient code without
overhead of object-based data frame representation. \Sys also supports domain-specific relational optimizations,
generalizing them for a case when program may contain both relational and non-relational operations.

\Sys implementation is based on the Julia programming language. 
\Sys leverages \hpat~\cite{HPAT} to automatically extract parallelism based on the semantics of the program, distribute data between nodes based
on heuristics about the data analytics domain and generate parallel code with highly efficient communication.
\Sys uses novel compilation techniques to implement data frame operations with minimal overhead and introduces domain-specific optimizations for relational operations.
We compare performance of \Sys with that of Python Pandas and Julia DataFrames, as examples of data frame packages, 
and \sparksql, as the most recent and best performing instance of a distributed big data system, demonstrating significant
performance advantages over both of these approaches.



This paper contributions are as follows:
\begin{itemize}

\item We present an end-to-end data analytics system that integrates relational data processing
with array computations in a scripting language using a productive data frames \api.
\item We describe domain-specific compiler techniques to automatically parallelize \df operations integrated in an array system.
\item We present novel compiler optimization techniques that provide the equivalent of \sql query optimization in a general compilation setting.
\item We demonstrate significant performance improvements over other frameworks that support relational data analytics in a general programming language.
For individual relational operations, \Sys outperforms both Python (Pandas) and \sparksql by 3.5x-177x.
For advanced analytics operations that require complex communication patterns,
\Sys is 1,000-20,000x faster than \sparksql.
Finally, for TPCx-BB Q25, Q26 benchmarks~\cite{Ghazal:2013} \Sys is 3-10x faster than \sparksql and provides good scalability.
\end{itemize}



\section{Background}\label{sec:back}

In this section, we provide an overview of data frames
and the existing approaches used by existing big data systems for data analytics.
In addition, we describe the \hpat and \pa systems, which we use as infrastructure for \Sys.

\subsection{Data Frames}\label{sec:dataframes}

Data analytics requires mathematical operations on arrays, as well as
relational operations on structured data. Hence, scripting languages such as
\Python, \R, and \Julia provide {\it data frame} abstractions~\cite{mckinneypandas,rdataframe,juliadataframe}.
A \df is a table or a two-dimensional array-like structure.
Columns in a \df are named and may have heterogeneous types but all the columns in a given \df have identical length.
In addition, columns of \dfs can be used in computation as regular arrays.
Data frames are similar to tables in a traditional relational database, but are designed to provide
high-level {\api}s such that domain experts (e.g. data
scientists and statisticians) can use them easily in procedural programs.

Data frames are typically implemented as libraries,
e.g. DataFrames.jl in Julia~\cite{juliadataframe} and Pandas in Python~\cite{mckinneypandas},
 that provide relational operations.
 These operations include {\it filtering} rows based on conditions,
 {\it joining} \dfs, and {\it aggregating} column values based on a key column.
Furthermore, they provide advanced analytics operations (e.g. moving averages)
and are tightly integrated with the underlying array system to
support array computations.

Data frames are essential for big data analytics systems. The popularity
of \dfs among data scientists confirms this fact:
Pandas in \Python and DataFrames.jl in \Julia are among the top popular
computational packages~\cite{pypulse,juliapulse}.
Furthermore, users of the DataFrame API of \sparksql increased by 153\%
in 2016~\cite{sparksurvey2016},
even though its interface is more restrictive than the \sql API.


\subsection{Big Data Systems}\label{sec:bigdatasys}
Current big data systems such as Apache Hadoop~\cite{hadoopweb} and Apache \spark~\cite{zaharia2010}
enable productive programming for clusters using the MapReduce paradigm~\cite{dean2008}.
The system provides high-level data-parallel operations such as \map
and \reduce, which are suitable for data processing,
but hides the details of parallel execution.
These systems are implemented as distributed runtime libraries, where a {\it master}
node schedules tasks on {\it slave} nodes.

However, these systems sacrifice performance for productivity and
are known to be orders of magnitude slower than
low-level hand-written parallel programs~\cite{jha2014tale}.
This is despite performance being the most important aspect for many users;
91\% of \spark users cited performance among the most important aspects for them
in a \spark survey, more than any other aspect~\cite{sparksurvey2015}.
Significant development effort has not found a solution (\spark has over 1000 contributors~\cite{sparksurvey2016})
since the problem is fundamental:
The distributed runtime library approach does not follow basic principles of parallel computing
such as avoiding sequential bottlenecks (the master node is inherently a sequential
bottleneck). Furthermore, the runtime task scheduling overhead is wasteful, since
most analytics programs can be statically parallelized~\cite{HPAT}.

Moreover, these systems are typically implemented in languages such as Java and
Scala that can have significant overheads~\cite{martin2015,david2016,ousterhout2015}.
The reason is that providing various data structures as part of \api
and implementing a complex distributed library is much easier in these object-oriented
languages. In addition, protections and facilities of a sandbox like
Java Virtual Machine (JVM) helps development and maintenance of these complex systems.
Hence, JVM overheads can be attributed to the distributed library approach,
but our compiler approach naturally avoids them due to code generation.

\subsection{\sparksql}\label{sec:sparksql}
Spark is a big data processing framework based on the Map-Reduce programming model~\cite{dean2008},
which is implemented as a master-slave distributed library.
It provides high-level operations such as \map and \reduce on linearly distributed collections
called Resilient Distributed Datasets (RDDs)~\cite{rdd2012}.
\sparksql is a \sql query processing engine built on top of \spark
that allows structured data processing inside \spark programs (as \sql strings)~\cite{armbrust2015}.
It compiles and optimizes \sql to Java byte code that
runs on top of RDD APIs for distributed execution.
\sparksql also provides data frame APIs in Python and Scala
that go through the same compiler pipeline.
However, these functional APIs are
restrictive for advanced analytics (our target domain).
For example, they cannot provide multiple aggregations, and are
restricted to simple expressions of data frame columns inside filter and aggregate operations.
In summary, \sparksql has various disadvantages for data scientists:
\begin{itemize}
\item It requires writing part of
the program in \sql which hurts productivity and is not type-safe.
\item It inherits the inefficiencies of distributed libraries
such as master-slave bottleneck and scheduling overheads.
\item It cannot provide parallel operations that do not fit in map-reduce paradigm
 such as moving averages efficiently (Section~\ref{sec:evaluation}).
 \end{itemize}

 In this paper, \sparksql is our baseline for distributed execution
 since it is the state-of-the-art distributed system that can support
 end-to-end data analytics programs.

\subsection{\hpat Overview}\label{sec:hpat:overview}
We build \Sys on top of High Performance Analytics Toolkit (\hpat), which
 demonstrated that it is possible to achieve productivity and
performance simultaneously for array computations in data analytics
and machine learning~\cite{HPAT}. \hpat
 performs static compilation of high-level scripting programs into high
performance parallel codes using domain-specific compiler techniques.
By generating scalable \mpic programs,
this approach enables taking advantage of compiler technologies (e.g. vectorization in C compilers),
as well as other HPC technologies (e.g. optimized collective communication routines of MPI).
In this work, we integrate \dfs with \hpat to create
a complete solution for data analytics that is both productive and efficient.
Automatic parallelization for distributed-memory machines is known to be a difficult problem~\cite{Kennedy:2007:RFH:1238844.1238851},
but \hpat can perform this task using a domain-specific data flow algorithm (which we extend).
Since \hpat avoids distributed library overheads such as runtime scheduling and master-slave coordination,
it is orders of magnitude faster than other systems such as \spark~\cite{zaharia2010}.
Furthermore, \hpat can generate parallel code to call existing HPC libraries such as HDF5~\cite{folk1999hdf5}, ScaLAPACK~\cite{choi1992scalapack}, and
\INTELR~DAAL~\cite{IntelDAAL}.
Section~\ref{sec:compiler} includes more details about the compilation pipeline of \hpat.

\hpat is built on top of \pa compiler infrastructure,
which is designed to extract parallel patterns from high-level Julia programs.
These patterns include \map, \reduce, \cartesianmap, and \stencil.
For example, \pa identifies array operations such as \CodeIn{-},
\CodeIn{!}, \CodeIn{log}, \CodeIn{exp}, \CodeIn{sin}, etc. as having \map semantics.
Then, \pa generates a common ``parallel for'' or \parfor representation that
allows a unified optimization framework for all the parallel patterns.

\subsection{Fault Tolerance}\label{sec:ft}
Fault tolerance is a major concern for large, unreliable clusters.
\spark provides fault tolerance using lineage of operations on RDDs, while \hpat provides
automatic minimal checkpoint/restart~\cite{dongarra2015fault}.
\Sys does not provide fault tolerance for failures during relational operations, since
we found the portion of our target programs with relational operations to be
significantly shorter than the mean time between failure (MTBF) of moderate-sized clusters.
Moreover, recent studies have shown that in practice most clusters
consist of 30-60 machines which is a scale at which fault tolerance is
not a big concern~\cite{ren2013}.
In essence, the superior performance of \Sys
helps users avoid paying the fault tolerance overheads by keeping execution times short
 (as Section~\ref{sec:evaluation} demonstrates).
On the other hand, iterative machine learning algorithms could require fault tolerance, which
\hpat provides.
\sparksql also does not provide fault tolerance for relational operations.

\section{\Sys Syntax}\label{sec:syntax}
%
%


\begin{table*}[ht!]
\footnotesize
\begin{minipage}{\linewidth}
\centering
  \begin{tabular}{| p{3.2cm} | p{4.3cm} | p{4.2cm} | p{4.5cm} |}
    \hline
    \textbf{Operations} & \textbf{\Julia} & \textbf{\sql} & \textbf{\Sys} \\
    \hline
    \hline
    \textbf{Projection} & \CodeInsmall{v = df[:id]}  & \CodeInsmall{select id} \newline \CodeInsmall{from t} & \CodeInsmall{v = df[:id]} \\
    \hline
    \textbf{Filter} &  \CodeInsmall{df2 = df[df[:id].<100,:]} & \CodeInsmall{select *} \newline \CodeInsmall{from table} \newline \CodeInsmall{where id<100} & \CodeInsmall{df2 = df[:id<100]}\\
    \hline
    \textbf{Join} & \CodeInsmall{rename!(df2,:id,:cid)} \newline \CodeInsmall{df3 = join(df1, df2, on=:id)} &
    \CodeInsmall{select *} \newline \CodeInsmall{from t1 join t2} \newline \CodeInsmall{on t1.id=t2.cid} & \CodeInsmall{df3 = join(df1, df2, :id==:cid)} \\
    \hline
    \textbf{Aggregate} & \CodeInsmall{df2 = by(df,:id, df ->} \newline \CodeInsmall{  DataFrame(} \newline \CodeInsmall{  xc = sum(df[:x].<1.0), } \newline \CodeInsmall{  ym = mean(df[:y])))}
                    &  \CodeInsmall{select count(case when x<1.0} \newline \CodeInsmall{  then 1 else null) as xc, } \newline \CodeInsmall{  avg(y) as ym } \newline \CodeInsmall{from t } \newline \CodeInsmall{group by id}
                    & \CodeInsmall{df2 = aggregate(df1, :id,} \newline \CodeInsmall{  :xc = sum(:x<1.0),} \newline \CodeInsmall{ :ym = mean(:y))}\\
    \hline
    \textbf{Concatenation} & \CodeInsmall{df3 = [df1; df2]} & \CodeInsmall{select * from t1 union all} \newline \CodeInsmall{select * from t2} & \CodeInsmall{df3 = [df1; df2]} \\
    \hline
    \textbf{Cumulative Sum} & \CodeInsmall{cumsum(df[:x])} & \CodeInsmall{select sum(x) over (rows}
           \newline \CodeInsmall{between unbounded preceding} \newline \CodeInsmall{and current row)} \newline \CodeInsmall{from t1} & \CodeInsmall{cumsum(df[:x])} \\
    \hline
    \textbf{Simple Moving Average \newline(SMA)} & \CodeInsmall{for i in 2:size(x,1)-1} \newline
        \CodeInsmall{ A[i] = (df[:x][i-1]+} \newline \CodeInsmall{\hspace{2ex}df[:x][i]+df[:x][i+1])/3.0} \newline \CodeInsmall{end} & \CodeInsmall{select avg(x) over (rows}
        \newline \CodeInsmall{between 1 preceding} \newline \CodeInsmall{and 1 following)} \newline \CodeInsmall{from t1} &
        \CodeInsmall{A = stencil(x->} \newline \CodeInsmall{(x[-1]+x[0]+x[1])/3.0,df[:x])} \\
    \hline
    \textbf{Weighted Moving Average \newline(WMA)} & \CodeInsmall{for i in 2:size(x,1)-1} \newline
        \CodeInsmall{ A[i] = (df[:x][i-1]+} \newline \CodeInsmall{\hspace{2ex}df[:x][i]+2*df[:x][i+1])/4.0} \newline \CodeInsmall{end} & \CodeInsmall{select (lag(x,1) over (rows}
        \newline \CodeInsmall{between 1 preceding} \newline \CodeInsmall{and 1 following) + 2*x + } \newline \CodeInsmall{lead(x,1) over (rows}
        \newline \CodeInsmall{between 1 preceding} \newline \CodeInsmall{and 1 following))/4.0} \newline \CodeInsmall{from t1} &
         \CodeInsmall{A = stencil(x->} \newline \CodeInsmall{(x[-1]+2*x[0]+x[1])/4.0,df[:x])} \\
    \hline
  \end{tabular}
\end{minipage}
\caption{Examples demonstrating relational and analytics API of Julia (DataFrames.jl), SQL and \Sys.}
\label{table:syntax}
\end{table*}

The goal of \Sys is to provide high-level \df abstractions that are
flexible, type-safe, and integrate seamlessly with array computations.
We make our APIs similar to \Juliadf to facilitate adoption, but
provide syntactic sugars based on the patterns we have observed in
data analytics programs.

\subsection{Data Frames API}


\paragraph{Input \dfs:}
To specify the schema and read a \df, we extend the \CodeIn{DataSource} construct of \hpat,
which is used for reading input data. For example, the following code reads a \df with three columns from an HDF5 file:

\begin{lstlisting}[language=Julia,escapeinside={(*}{*)},mathescape,numbers=none]
df = DataSource(DataFrame{:id=Int64, :x=Float64,
         :y=Float64}, HDF5, "/data/data.hdf5")
\end{lstlisting}

The first argument is the schema of the \df. Similar to DataFrames.jl, Julia's symbols (e.g. \CodeIn{:id})
are used for referring to column names. Each column's type is also specified.
The equivalent code in \spark (Python) follows:

\begin{lstlisting}[language=Python,escapeinside={(*}{*)},mathescape,numbers=none]
schema = StructType(\
             [StructField("id",LongType(),True),\
              StructField("x",DoubleType(),True),\
              StructField("y",DoubleType(),True)])
df = spark.createDataFrame(data, schema)
\end{lstlisting}




\paragraph{Projection:}
One can use columns of \Sys as arrays which is equivalent to the {\it projection}
relational operation (see example in second row of Table~\ref{table:syntax}).

\paragraph{Filter:}
\Sys allows filtering data frames using a conditional expression
as shown in the third row of Table~\ref{table:syntax}.
This example filters all the row whose
  ``id'' column value is less than 100.
For convenience, the user can refer to columns by just their names and use simple mathematical
operators instead of element-wise operators (see desugaring in Section~\ref{sec:desugar}).
However, any array expression that results in a boolean array can be used,
and referring to any array in the program (including columns of other \dfs) is allowed.

%
%

\paragraph{Join:}
\Sys provides the \join operation as shown in the fourth row of Table~\ref{table:syntax}.
Note that unlike \Juliadf, our API allows different column names as keys for the two input tables.

%

\paragraph{Aggregate:} \Sys provides ``split-and-combine'' operations through a flexible \CodeIn{aggregate()} syntax,
which is demonstrated in the fifth row of Table~\ref{table:syntax}.
Instead of the anonymous lambda syntax of \Juliadf, we extend
the \CodeIn{aggregate()} call to accept column assignment expressions shown as syntactic sugar.


\paragraph{Concatenation:} \Sys provides vertical concatenation of \dfs with the same schema,
demonstrated in the fifth row of Table~\ref{table:syntax}. 
%
%
%

%
%

\paragraph{Cumulative sum:}
Cumulative sum (\CodeIn{cumsum}) calculates the sequence of partial sums over an array. It is
an example of built-in analytics functions of scripting languages \Sys provides
(sixth row of Table~\ref{table:syntax}). In \sql, it requires defining a window
from the first row of the table to the current row being processed.

\paragraph{Simple Moving Average (SMA):}
Simple Moving Average (SMA) is a {\it data smoothing} technique where for each value an average
using neighboring values is calculated (sixth row of Table~\ref{table:syntax}).
Julia does not provide a specific syntax for SMA;
Julia users typically write these operations as for loops since Julia compiles
loops to native code.
However, Python (Pandas) provides SMA using {\it rolling} windows:
\begin{lstlisting}[language=Python,escapeinside={(*}{*)},mathescape,numbers=none]
A = df1['x'].rolling(3,center=True).mean()
\end{lstlisting}
\sql requires defining a window and using the built-in \CodeIn{avg()} function.
\Sys provides one-dimensional stencil API to support moving averages,
which improves productivity and allows parallelization.

\paragraph{Weighted Moving Average (WMA):}
Weighted Moving Average (WMA) is similar to SMA, except that the user provides
the weights for the average operation. Again, Julia requires a loop, while \Sys handles WMA
using stencils.
Python (Pandas) provides WMA by accepting a user lambda for rolling windows:
\begin{lstlisting}[language=Python,escapeinside={(*}{*)},mathescape,numbers=none]
A = df1['x'].rolling(3,center=True).
    apply(lambda x: (x[0]+2*x[1]+x[2])/4)
\end{lstlisting}
\sql provides \CodeIn{lag()} and \CodeIn{lead()} functions to
access neighboring rows in a window using relative indices.

\subsection{Example}\label{sec:fullexample}

Consider the following example data analytics program written using \Sys,
which is inspired by TPCx-BB Q26 benchmark~\cite{Ghazal:2013}:

\begin{lstlisting}[language=Julia,escapeinside={(*}{*)},mathescape,numbers=none,
    linebackgroundcolor={\ifnum \value{lstnumber}=3 \color{gray!30}
      \else \ifnum \value{lstnumber}=4 \color{gray!30}
      \else \ifnum \value{lstnumber}=7 \color{gray!30}
      \else \ifnum \value{lstnumber}=8 \color{gray!30}
      \else \ifnum \value{lstnumber}=14 \color{gray!30}
      \else \ifnum \value{lstnumber}=17 \color{gray!30}
      \else \ifnum \value{lstnumber}=18 \color{gray!30}
      \else \ifnum \value{lstnumber}=19 \color{gray!30}
      \else \ifnum \value{lstnumber}=21 \color{gray!30}
      \fi \fi \fi \fi \fi \fi \fi \fi \fi},  linebackgroundsep=0.39em, linebackgroundwidth=33em
  ]
@acc hiframes function customer_model(min_count,
     num_centroids, iterations, file_name)
 store_sales = DataSource(DataFrame{:s_item_sk=Int64,
     :s_customer_sk=Int64}, HDF5, file_name)
 item = DataSource(DataFrame{:i_item_sk=Int64,
     :i_class_id=Int64}, HDF5, file_name)
 sale_items = join(store_sales, item,
     :s_item_sk==:i_item_sk)
 c_i_points = aggregate(sale_items, :s_customer_sk,
    :c_i_count = length(:s_item_sk),
    :id1 = sum(:i_class_id==1),
    :id2 = sum(:i_class_id==2),
    :id3 = sum(:i_class_id==3))
 c_i_points = c_i_points[:c_i_count>min_count]
 c_i_points[:id3] = (c_i_points[:id3]-
    mean(c_i_points[:id3]))/var(c_i_points[:id3])
 samples = transpose(typed_hcat(Float64,
    c_i_points[:c_i_count], c_i_points[:id1],
    c_i_points[:id2], c_i_points[:id3])
 model = HPAT.Kmeans(samples, num_centroids, iterations)
 return model
end
\end{lstlisting}

This program performs {\it market segmentation} where it builds a model of separation
 for customers based on their purchase behavior\footnote{\url{http://www.tpc.org/tpc_documents_current_versions/pdf/tpcx-bb_v1.1.0.pdf}}.
It reads \CodeIn{store\_sales} and \CodeIn{item} \dfs from file and
joins them. Then, it forms training features based on the number of items each customer
bought in total and in different classes. The program also filters the customers that bought less than
a minimum number. {\it Feature scaling} is used for column \CodeIn{:id3} based on its mean and variance (\CodeIn{var}).
The next step is {\it matrix assembly} where the training matrix is formed: The call \CodeIn{typed\_hcat}
is a standard Julia operation where arrays are concatenated horizontally (including type conversion).
Also, the matrix is transposed
since Julia has column major layout and features need to be on the same column.
Finally, K-means clustering algorithm is called to train the model.
Note that this program is simplified and there could be much more mathematical
operations (array computations) such as data transformations and feature scaling. Furthermore,
the user might write a custom machine learning algorithm instead of calling a library. 
The equivalent \sparksql version is about 2$\times$ longer and includes a \sql
 string for relational operations.
\section{Compiling Data Frames}\label{sec:compiler}

\begin{figure*}[t]
     \centering
     \def\svgwidth{\textwidth}
     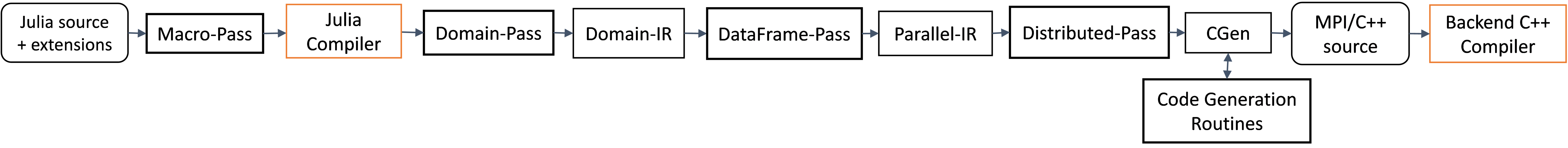
     \caption{Compiler pipeline of \Sys. For \df support, we added
     \dfpass and extended \CodeIn{Macro-Pass}, \CodeIn{Domain-Pass},
     \CodeIn{Distributed-Pass}, and \CodeIn{CGen} routines.}
     \label{fig:pipeline}
\end{figure*}

In this section, we describe \Sys's compiler implementation.
\Sys uses a novel {\it dual representation} approach: all
columns are individual arrays in the AST which allows Julia and \hpat
to optimize the program. However, \Sys uses \df metadata when necessary for
relational transformations and optimizations.
Figure ~\ref{fig:example1} shows the initial code of a running example to illustrate the transformations of
the \Sys compiler pipeline.

\begin{figure}
\begin{lstlisting}[language=Julia,escapeinside={(*}{*)},mathescape,numbers=none]
df = DataSource(DataFrame{:id=Int64, :x=Float64,
    HDF5, "data.hdf5")
df2 = aggregate(df, :id, :c = sum(:x<1.0))
\end{lstlisting}
\vspace{-1ex}
\caption{Running example \Sys source code.}
\vspace{-1ex}
\label{fig:example1}
\end{figure}

\begin{figure}
\begin{lstlisting}[language=Julia,escapeinside={(*}{*)},mathescape,numbers=none]
Expr(:meta, (:df->((:id,Int64), (:x,Float64)), :df2->((:id,Int64), (:c,t1))))
_df_id = DataSource(Int64, "id", HDF5, "data.hdf5")
_df_x = DataSource(Int64, "x", HDF5, "data.hdf5")
expr_arr1 = _df_x .< 1.0
t1 = typeof(sum(expr_arr1))
arg_arrs_in = [_df_id, _df_x]
arg_arrs_out = HiFrames.API.aggregate(arg_arrs_in, :id, [(expr_arr1, sum)])
_df2_id::Vector{Int64} = arg_arrs_out[1]
_df2_c::Vector{t1} = arg_arrs_out[2]
\end{lstlisting}
\vspace{-1ex}
\caption{Running example after Macro-Pass.}
\vspace{-1ex}
\label{fig:example2}
\end{figure}

\begin{figure}
\begin{lstlisting}[language=Julia,escapeinside={(*}{*)},mathescape,numbers=none]
size1 = get_h5_size("id", "data.hdf5")
_df_id = alloc(Int64, size1)
_df_x = alloc(Float64, size1)
h5read(_df_id, "id", "data.hdf5", size1)
h5read(_df_x, "x", "data.hdf5", size1)
expr_arr1 = map(.<, _df_x)
Expr(:aggregate, :df, :df2, (expr_arr1, sum))
\end{lstlisting}
\vspace{-1ex}
\caption{Running example after Domain-Pass.}
\vspace{-1ex}
\label{fig:example3}
\end{figure}

\begin{figure}
\begin{lstlisting}[language=C,escapeinside={(*}{*)},mathescape,numbers=none,morekeywords={int64_t}]
// allocations, read
MPI_Comm_size(MPI_COMM_WORLD, &npes);
H5Dopen2(...);
size1 = H5Sget_simple_extent_ndims(space_id_2);
_df_id = new int64_t[size1/npes];
H5Sselect_hyperslab(...);
H5Dread(...);
// same alloc, read calls for _df_x
expr_arr1 = new bool[size1/npes];
for(int i=0; i<size1/npes; i++) {
  expr_arr1[i] = (_df_x[i]<1.0);
}
for(int i=0; i<size1/npes; i++) {
  int pe_id = _df_id[i]%npes;
  send_count[pe_id]++;
}
MPI_Alltoall(send_count,...);
for(int i=0; i<size1/npes; i++) {
  // pack data in buffers for different processors
}
MPI_Alltoallv(temp_df_id,send_count, recv_df_id,...);
MPI_Alltoallv(temp_df_x,send_count, recv_df_x,...);
for(int i=0; i<recv_df_id.size(); i++) {
   // aggregate using hash table
   int64_t key = recv_df_id[i];
   int write_ind = agg1_table[key];
   _df2_c[write_ind] += expr_arr1[i];
}
\end{lstlisting}
\vspace{-1ex}
\caption{Running example output C code.}
\label{fig:example4}
\vspace{-1ex}
\end{figure}

Figure~\ref{fig:pipeline} provides an overview of the compiler pipeline.
For \df support, we added
\dfpass and extended \CodeIn{Macro-Pass}, \CodeIn{Domain-Pass},
and \CodeIn{Distributed-Pass}. We also added code generation routines
for relational operations in \CodeIn{CGen}.

\subsection{Macro-Pass}\label{sec:desugar}
\CodeIn{Macro-Pass} is called at the macro stage and is responsible for desugaring
\Sys operations to make sure Julia can compile the program.
In addition, the types of all variables should be available to the Julia compiler
for complete type inference.
Here, we desugar \df operations into
regular array operations and function calls, and annotate variables with types using
domain knowledge.
In general, each \df column is a regular array in the AST, but \df metadata
is included to enable relational operations and optimizations.
The output of the \CodeIn{Macro-Pass} of our running example is shown in Figure ~\ref{fig:example2}.

\paragraph{Input \dfs:}
\Sys desugars read operations for \dfs into separate \CodeIn{DataSource()} calls
for each column:

\begin{lstlisting}[language=Julia,escapeinside={(*}{*)},mathescape,numbers=none]
df_name = DataSource(DataFrame{:c1=<c1_type>,
    :c2=<c2_type>, ...}, f_typ, file_path)
\end{lstlisting}
\begin{lstlisting}[language=Julia,escapeinside={(*}{*)},mathescape,numbers=none]
# assert length(_df_c1)==length(_df_c2)
Expr(:meta, (:df->((:id,Int64), (:x,Float64)))
_df_c1 = DataSource(<c1_type>, "c1", f_typ,file_path)
_df_c2 = DataSource(<c2_type>, "c2", f_typ,file_path)
...
\end{lstlisting}

Each column is an array but
various metadata for the \df is inserted in the metadata section of the
AST (\CodeIn{Expr(:meta)} node in Julia).
In addition, columns of the \df are set to have the same length,
which enables many array optimizations such as fusion.
Furthermore, \df column references are desugared to the underlying array (\CodeIn{df[:id]} to \CodeIn{\_df\_id}).

\paragraph{Filter:}
\Sys desugars filter operations into regular function calls on
arrays.
Since the number of columns of \dfs is variable, \Sys packs the columns
into an array of arrays as follows:

\begin{lstlisting}[language=Julia,escapeinside={(*}{*)},mathescape,numbers=none]
df2 = df1[e(:c1, ...)]
\end{lstlisting}
\begin{lstlisting}[language=Julia,escapeinside={(*}{*)},mathescape,numbers=none]
Expr(:meta, (:df1->((:c1,T1), (:c2,T2)...), :df2->((:c1,T1), (:c2,T2)...)))
# replace column references of e to underlying arrays
e = ast_walk(replace_column_refs, e)
# convert scalar operations of e to element-wise operations
e = ast_walk(replace_opr_vector, e)
expr_arr = e
# n is number of columns of df1 which is constant
arg_arrs_in = Array(Array, n)
arg_arrs_in[1] = _df1_c1
arg_arrs_in[2] = _df1_c2
...
arg_arrs_out = HiFrames.API.filter(expr_arr, arg_arr_in)
_df2_c1::Vector{T1} = arg_arrs_out[1]
_df2_c2::Vector{T2} = arg_arrs_out[2]
...
\end{lstlisting}

For the translation of relational operations such as filter,
\Sys reads metadata of the involved \df from metadata.
For example, the types of output columns are assigned using
the metadata information available from the input \df.
This desugaring method ensures that Julia can compile the generated code and perform complete type
inference.

\paragraph{Join:} Join desugaring is similar to filter, except
that there are two inputs \dfs and the join expression should be specified.
Currently, we support inner join on equal keys but relaxing this limitation is straightforward.

% ...

\paragraph{Aggregate:} 
Similar to filter, aggregate expressions are
translated to replace scalar operations with element-wise counterparts and
to replace column references with underlying arrays. However,
the type of the output columns cannot be determined at the macro stage easily.
Therefore, we generate dummy calls that apply the reduction functions on expression arrays
to find the output type. Furthermore, for each output column, the expression array and the
reduction function, which form a tuple, are passed as inputs to the aggregate function call.
The aggregation key is also passed as input.

\begin{lstlisting}[language=Julia,escapeinside={(*}{*)},mathescape,numbers=none]
df2 = aggregate(df1, :c1, :c4=f1(e1(:c2,...)),
                          :c5=f2(e2(:c3,...)),
                          ...)
\end{lstlisting}
\begin{lstlisting}[language=Julia,escapeinside={(*}{*)},mathescape,numbers=none]
Expr(:meta, (:df1->((:c1,T1), (:c2,T2)...), :df2->((:c1,t1), (:c4,t2)...)))
# replace column references of e1,e2,... to underlying arrays
# convert scalar operations of e1,e2,... to element-wise operations
expr_arr1 = e1
expr_arr2 = e2
...
# dummy call to get type of f1 output on expr_arr1
t1 = typeof(f1(expr_arr1))
t2 = typeof(f2(expr_arr2))
...
arg_arrs_out = HiFrames.API.aggregate(:c1, arg_arrs_in, [(expr_arr1, f1), (expr_arr2, f2), ...])
_df2_c1::Vector{t1} = arg_arrs_out[1]
_df2_c4::Vector{t2} = arg_arrs_out[2]
...
\end{lstlisting}

\paragraph{Concatenation:} \Sys desugars union of \dfs into vertical concatenation
of columns (Julia's \CodeIn{vcat()} call)
after making sure schemas are equal.

% ...

\subsection{Domain-Pass}

Julia then translates the code to its internal representation and performs type inference.
Next, the \CodeIn{Domain-Pass} encapsulates relational operations into their own AST nodes
so that \hpat and \pa can be applied.
The output of the \CodeIn{Domain-Pass} of our running example is shown in Figure ~\ref{fig:example3}.

\begin{lstlisting}[language=Julia,escapeinside={(*}{*)},mathescape,numbers=none]
Expr(:filter, expr_arr, out_df, in_df,...)
Expr(:join, key_arrs, out_df, in_df1, in_df2,...)
Expr(:aggregate, key_arr, out_df, in_df, expr_arrs, reduction_funcs,...)
\end{lstlisting}

\CodeIn{Domain-Pass} also simplifies the AST by removing all the unnecessary code
generated after Julia compilation. For example, we remove the array of arrays variables
used for passing data frames and the related packing/unpacking code generated.

Since \Sys transforms relational operations into fully-fledged AST nodes,
the optimizations of \pa and \hpat can transparently work with relational operations as well.
For example, \pa dead code elimination will remove unused columns
({\it column pruning})
using the knowledge of the whole program, while \sparksql performs {\it column pruning} only within the \sql context.

Moreover, \CodeIn{Domain-Pass} can perform pattern matching for common
patterns of analytics workloads (before the structure is lost in later passes).
For example, we match the \CodeIn{transpose(typed\_hcat())} pattern,
since it is used for machine learning matrix assembly (see example of Section~\ref{sec:fullexample}).
\Sys replaces the original code with a call to \CodeIn{HiFrames.API.transpose\_hcat()}, which
has an optimized code generation routine in the backend (\Sys extension of \CodeIn{CGen}).
We found this optimization to be significant for this step (not presented in this paper).

After \CodeIn{Domain-Pass}, we call the \CodeIn{Domain-IR} pass of \pa, which is
responsible for normalizing the AST for further analysis. We insert
the \dfpass after \CodeIn{Domain-IR} since some analyses,
such as liveness analysis, are only available after this normalization.

\subsection{\dfpass: Relational Optimizations}
\dfpass is responsible for optimizing relational operations.
Relational databases and other \sql systems such as \sparksql
usually optimize queries using a query tree and applying various
rule-based transformations repeatedly.
However, \Sys receives some of the optimizations implemented in \sql
systems for ``free'' by design. For example, the Julia compiler performs constant folding and
common subexpression elimination and there is no need for \Sys to implement them.
In addition, \pa performs advanced optimizations such as loop fusion and intermediate array elimination.
However, some optimizations are specific to relational operations and
are not handled by general compilers.
Performing these optimizations is challenging in a general program AST since relational operations of \Sys
are spread across the AST. For example, there could be array computation
or sequential code between two relational nodes that need to be transformed.

We address this challenge in \dfpass by using the following heuristic-based approach.
Similar to traditional databases, \dfpass starts by constructing a query tree of operations.
However, unlike databases, this tree includes only relational operations while other nodes in the AST are ignored at this stage.
The root node of the tree is the output data frame and each internal node corresponds to a relational operator.
The leaf nodes represent the input data frames.
Similar to databases, \dfpass then traverses the tree and checks the rules to find
the transformations that can be applied.
However, we need to make sure a transformation is valid for the full program before applying it
since the query tree is only a partial view.
For example, a column of a \df could be used in array computations between
 two relational operations, and their
transformation could change the result of the computation.
To make sure a transformation does not change the semantics of the program,
we use liveness analysis to find and inspect potential references to the columns of the involved data frames in any
node that can be executed between the involved operators.
Currently, we perform relational transformations within basic blocks only, since
our use cases do not require transformations across control flow. Extending this pass to
handle control flow is straightforward (e.g. make sure the earlier relational node is
in a dominant block with respect to the later node).

\begin{figure}[ht!]
    \centering
    \begin{subfigure}{\linewidth}
    \begin{lstlisting}[language=Julia,escapeinside={(*}{*)},mathescape,numbers=none]
# customer dataframe has columns :id, :phone
# order dataframe has columns :customerId, :amount
cust_ord = join(customer, order, :id==:customerId)
...
cust_ord = cust_ord[:amount>100.0]
    \end{lstlisting}
    \vspace{-1ex}
    \begin{lstlisting}[language=Julia,escapeinside={(*}{*)},mathescape,numbers=none]
# is transformed to the equivalent of:
order2 = order[:amount>100.0]
...
cust_ord = join(customer, order2, :id==:customerId)
    \end{lstlisting}
    \caption{\label{fig:pushdowncode}}
    \end{subfigure}
    \begin{subfigure}{.4\linewidth}
      \includegraphics[scale=0.3]{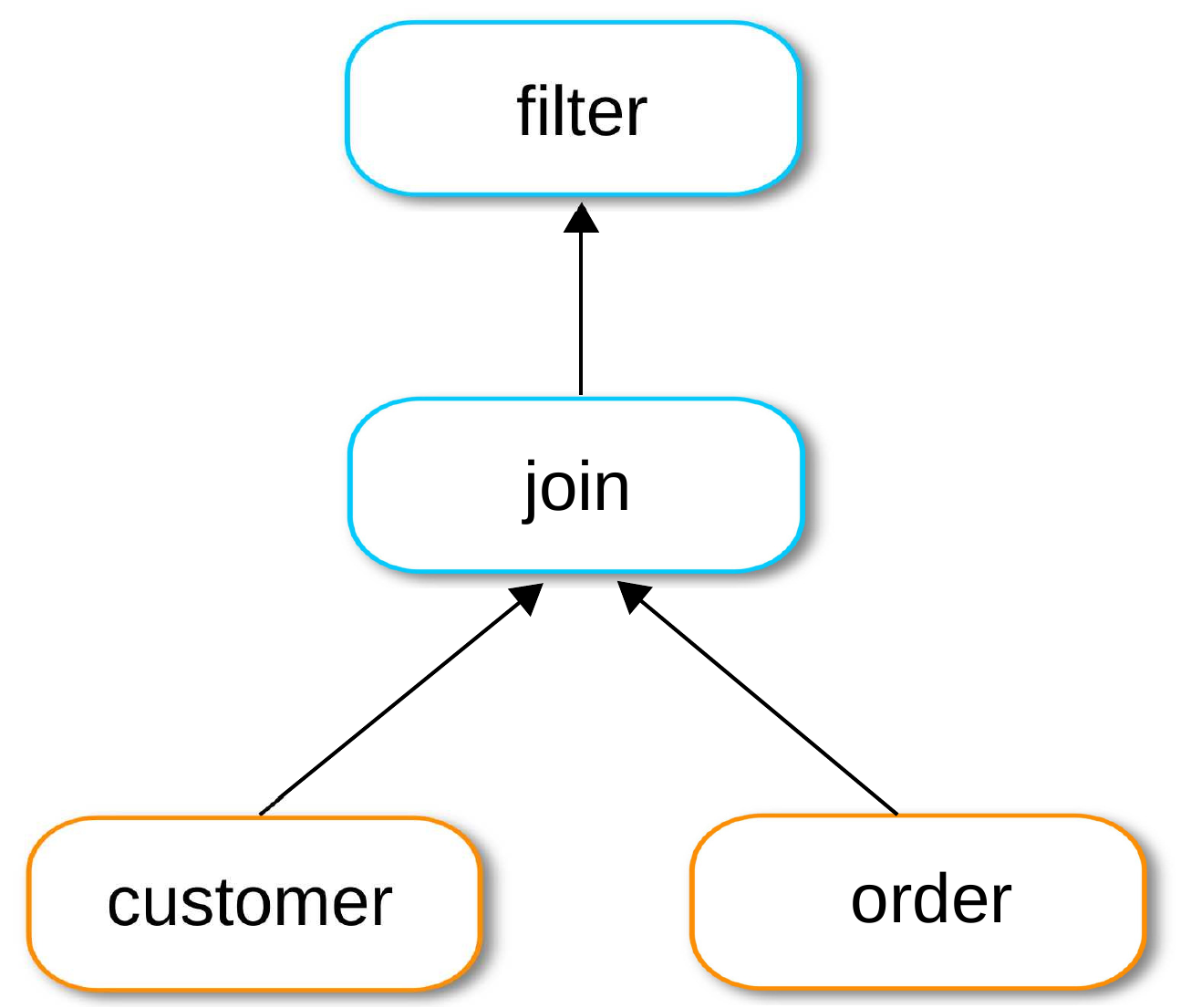}
      \caption{\label{fig:unoptimizedtree}}
    \end{subfigure}
    \hskip2em
    \begin{subfigure}{.4\linewidth}
      \includegraphics[scale=0.3]{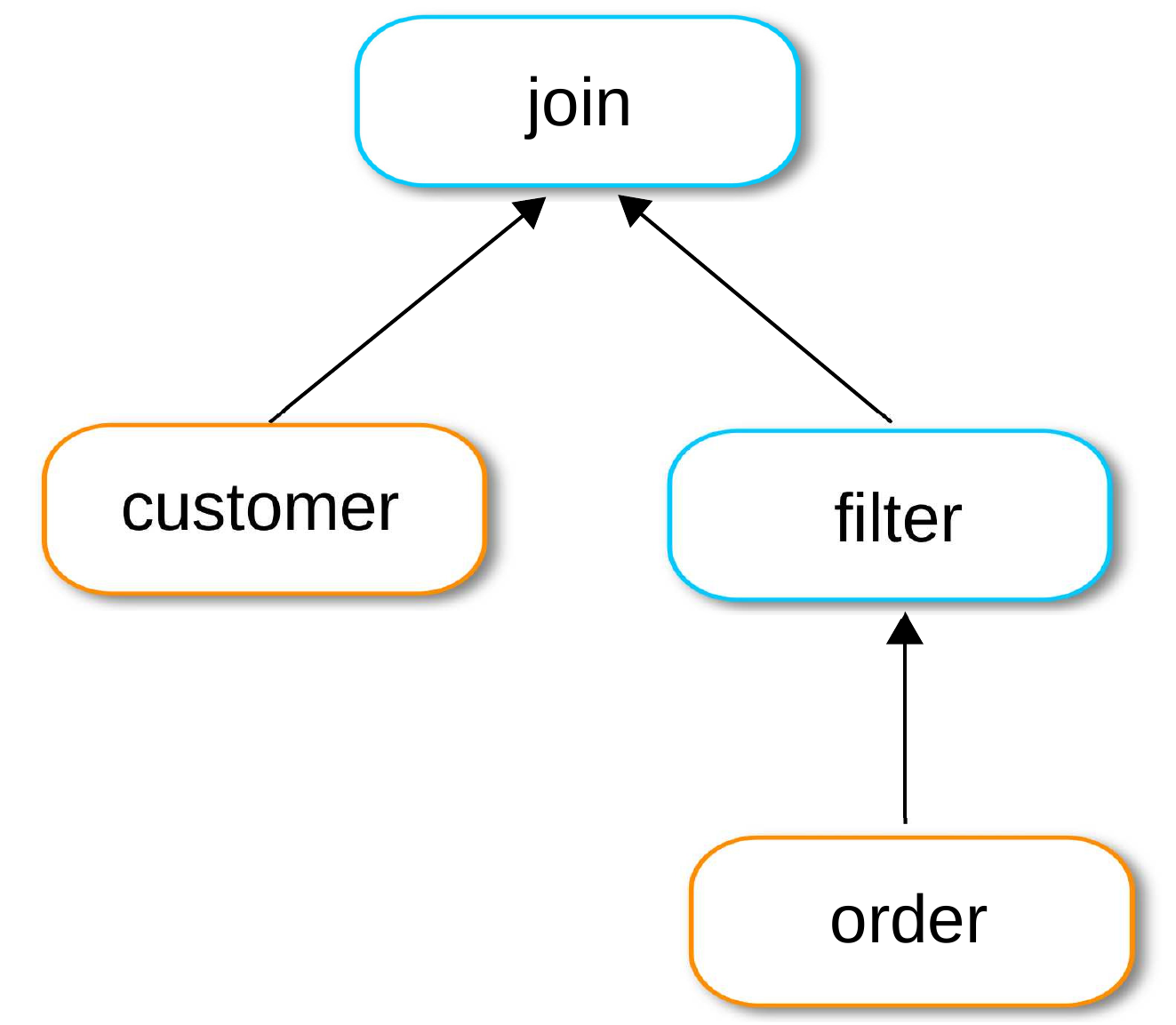}
      \caption{\label{fig:optimizedtree}}
    \end{subfigure}
    \vspace{-1ex}
    \caption{\textbf{(a)} Example program before and after {\it push predicate through join}
    transformation \textbf{(b)} Unoptimized tree \textbf{(c)} Optimized tree}
    \label{fig:pushdown}
    \vspace{-2ex}
\end{figure}

We use this approach to implement the {\it push predicate through join}~\cite{pushdown1993} optimization,
which we found to be the most important for our current workloads.
This optimization is potentially applicable when the output table of a join
operation is filtered based only on the attributes of an input table.
In this case, the input table can be filtered instead, which can reduce
the cost of the join operation substantially by decreasing the data size.
Figure~\ref{fig:pushdown} illustrates an example program before and after
the transformation (\ref{fig:pushdowncode}), and the corresponding
before and after trees (\ref{fig:unoptimizedtree} and \ref{fig:optimizedtree}).

After \dfpass, \CodeIn{Parallel-IR} is called which lowers
computations into \parfor nodes and performs more optimizations such as loop fusion.

\subsection{Distributed-Pass}\label{sec:dist-pass}
Parallelization for distributed-memory architectures is performed in \CodeIn{Distributed-Pass}.
The first step is distribution analysis for arrays and \parfor nodes
to determine which arrays and computations should be parallelized.
\hpat uses a heuristic-based data flow approach where distribution methods form a meet-semilattice.
In a fixed-point iteration algorithm, the distribution method of each array and \parfor is updated
using domain-specific inference rules (i.e. transfer functions) for each node in the AST.
The default distribution (top element of the semilattice) is
one-dimensional block distribution (\CodeIn{1D\_BLOCK}),
which means all processors have
equal chunks of data except possibly the last processor.
However, the distribution can change all the way to replication (\CodeIn{REP}).

We extend the inference rules to support relational operations. Similar to
other inference rules of \hpat, we use domain knowledge for developing these
new rules. For example, all input and output arrays of an aggregate operation
should be replicated if any of them is replicated, which makes the aggregate operation
sequential. This is essentially assigning the meet of the distributions of the arrays
to all of them.

However, the output arrays of relational operations need extra attention. Even though
the input arrays could have one-dimensional block distribution, the output chunks can have variable
length since the output size is data dependent.
For example, filtering a data frame in parallel could result in different number
of rows on different processors.
This is an issue since some operations such as the distributed machine learning
algorithms \hpat provides require \CodeIn{1D\_BLOCK} distribution for their input arrays.
One could rebalance the data frames after every relational operation but
this can be very costly. The best approach is to rebalance only when necessary,
which we achieve using a novel technique.

\begin{figure}
    \centering
    \includegraphics[scale=0.4]{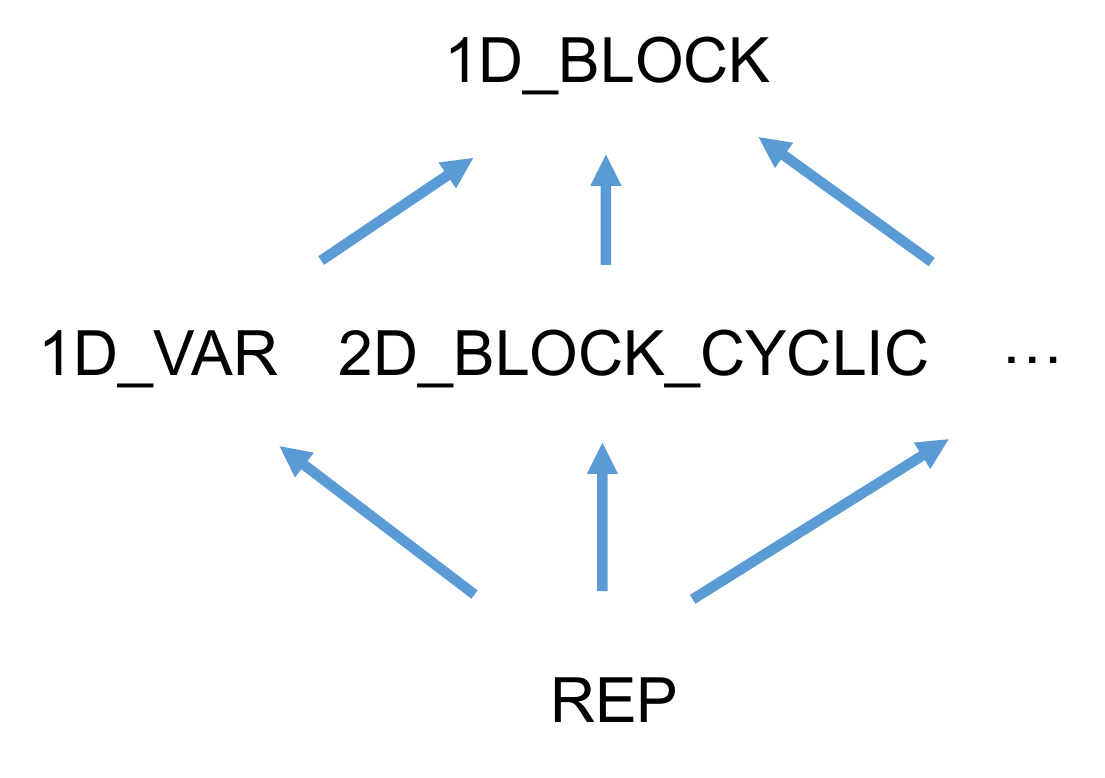}
    \caption{\Sys extends the meet-semilattice of \hpat distributions with \CodeIn{1D\_VAR}
    to support relational operations.}
    \label{fig:semilattice}
\end{figure}

To achieve this, we extend the meet-semilattice of
distributions with a new one-dimensional variable length (\CodeIn{1D\_VAR}) distribution.
Figure~\ref{fig:semilattice} illustrates the new meet-semilattice.
The transfer functions for output arrays of relational operations could be written as follows:

\noindent
$dist[out\_arrs] = \CodeIn{1D\_VAR} \wedge dist[in\_arr1] \wedge dist[in\_arr2] ...$

Furthermore, we allow \CodeIn{1D\_VAR} as input to operations
that require \CodeIn{1D\_BLOCK} during analysis. However, we generate a
rebalance call in the AST right before these operations.

This technique allows having
multiple parallel patterns in a program simultaneously
and takes full advantage of
 the meet-semilattice of Figure~\ref{fig:semilattice}. For example,
one could use linear algebra algorithms that require two-dimentional block cyclic distribution
along with relational operations, and generate data distribution conversions only when necessary.
This is not possible with distributed libraries like \spark and Hadoop, since they have
one-dimensional partitioning hard-coded in the system.
Evaluation of this feature is left for future work.

\subsection{Backend Code Generation in CGen}
We extend \CodeIn{CGen} with code generation routines for relational operations.
The output of this pass for our running example is shown in Figure ~\ref{fig:example4}.
Operations such as filter do not require communication by taking advantage of our
\CodeIn{1D\_VAR} distribution approach (Section~\ref{sec:dist-pass}).
On the other hand, aggregate and join require data shuffling since rows with the same
key need to be on the same processor for these operations (currently using
hash partitioning). We use \CodeIn{MPI\_Alltoallv()} collective communication
routine of MPI to perform data shuffling. However, since MPI requires the amount of data to be
known for each call, we use an initial \CodeIn{MPI\_Alltoall()} operation
for processors to coordinate the number of elements that will be communicated.
Optimizing and tuning communication operations is left for future work.
After data shuffling, join and aggregate operations are performed using
standard algorithms. We use {\it sort-merge} for join, with Timsort~\cite{peters2002timsort}
as the sorting algorithm. Aggregation is done using a hash table.

We add code generation routines for analytics operations as well.
For example, \CodeIn{cumsum} generates loops for local partial sums
and \CodeIn{MPI\_Exscan} for the required parallel scan communication.
Furthermore, stencils of \Sys generate near neighbor communication
and the associated border handling. To overlap communication and computation,
non-blocking communication is used (\CodeIn{MPI\_Isend}, \CodeIn{MPI\_Irecv} and \CodeIn{MPI\_Wait}).

Being able to take advantage of HPC technologies such as MPI
collective communication routines give a significant advantage to our approach.
This helps taking advantage of decades of development in the HPC domain.
On the other hand, the \sparksql approach requires building components
such as shuffle collective on top of a distributed library like Spark, which
can have significant overheads, increases development cost,
and increases system complexity substantially.

\section{Evaluation}\label{sec:evaluation}
In this section, we evaluate and compare the performance of analytics workloads on
 \sparksql and \Sys.
While \spark is a highly optimized production system
with over 1000 contributors~\cite{sparksurvey2016},
\Sys is currently a research prototype without significant performance tuning effort.
Nevertheless, performance comparison provides valuable insight about the two
approaches.

We use a local 4-node cluster for evaluation (144 total cores).
Each node has two sockets, each equipped with an \INTELR \XEONR E5-2699 v3
 processor (18-core ``Haswell'' architecture). The memory capacity per node is 128GB.
The nodes are connected through an Infiniband network.
We use Spark 2.0.1 for comparison which is the latest release at the time of
this study. It is configured to take advantage of all of system memory and
the Infiniband network.
The cluster software includes Julia 0.5 compiled with \INTELR \cpp Compiler 17,
\INTELR MPI 2017 configured with DAPL communication stack, and Python 3.5.2 (Anaconda 4.1.1 distribution).

\begin{figure*}[t]
\centering
\begin{subfigure}{.48\linewidth}
    \centering
    \def\svgwidth{\columnwidth}
    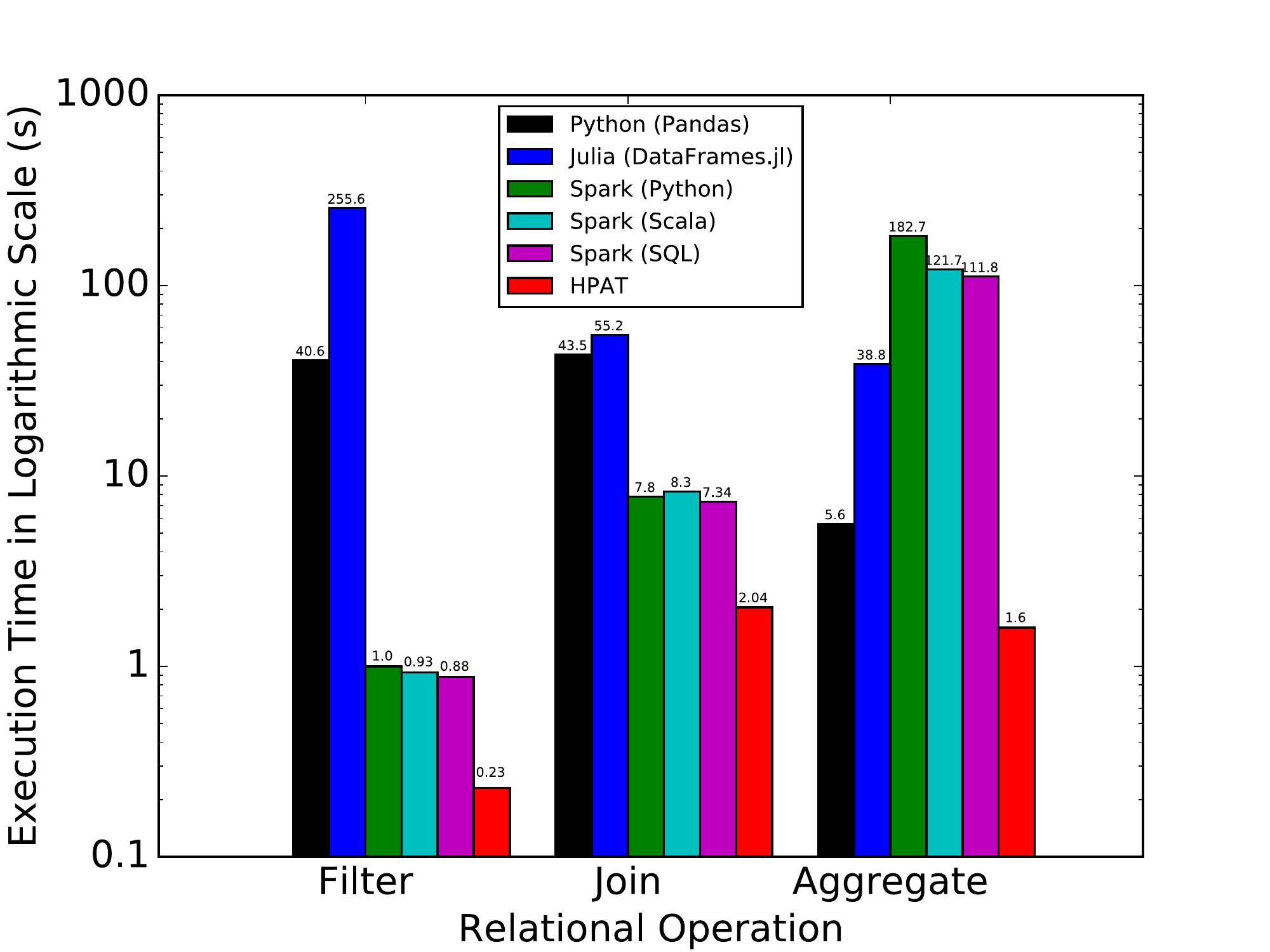
    \caption{Relational Operations (\Sys is 3.6$\times$-70$\times$ faster than \sparksql)}
    \label{fig:basic}
\end{subfigure}
\begin{subfigure}{.48\linewidth}
    \def\svgwidth{\columnwidth}
    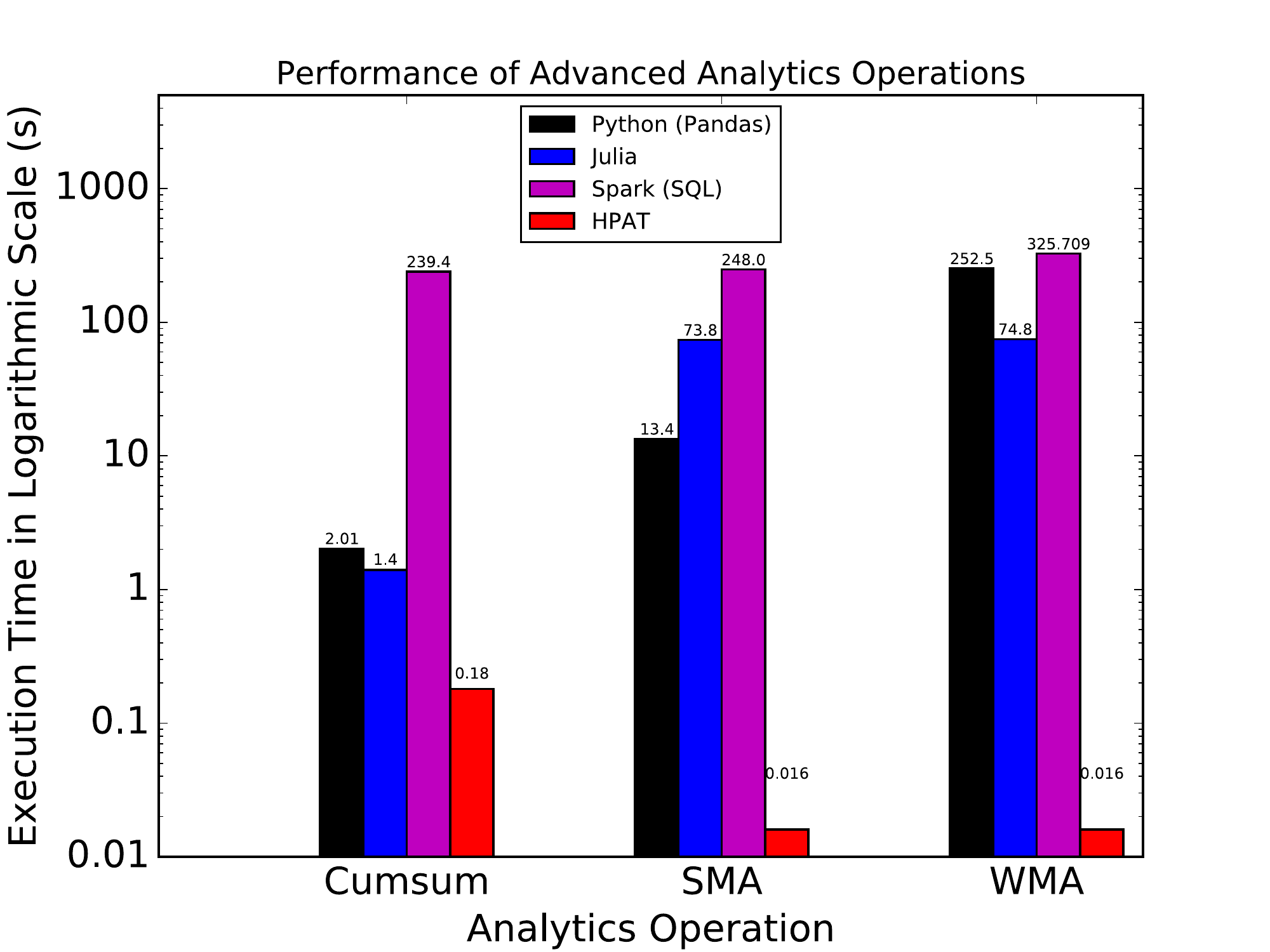
    \caption{Advanced Analytics Operations (\Sys is 11$\times$-15781$\times$ faster than \sparksql)}
    \label{fig:advanced}
\end{subfigure}
\caption{Performance of relational and analytics operations in various systems. Please note the
     logarithmic scale.}
\end{figure*}

\paragraph{Basic Relational Operations:}
We compare the performance of various systems that provide \df \api for the basic relational operations: filter,
join, and aggregate. The input tables
have an integer key field and two floating point numbers. The datasets are randomly
generated from uniform distribution to avoid load balance issues.
Dataset sizes are large enough to demonstrate parallel execution
while making sure sequential systems (Python and Julia)
do not run out of memory.
Input tables to filter, join, and aggregate operations have 2 billion, 0.5 million, and
256 million rows, respectively.

Figure~\ref{fig:basic} demonstrates that \Sys is 177$\times$, 21$\times$,
and 3.5$\times$ faster than Python (Pandas)
for filter, join, and aggregate, respectively. Furthermore, \Sys is 3.8$\times$,
3.6$\times$, and 70$\times$ faster than \sparksql.
The choice of Spark interfaces
does not affect our comparison significantly for these benchmarks.
For these experiments, \sparksql is expected to perform its best since
the cluster is small and the master-slave bottleneck does not affect performance
significantly. Furthermore, the benchmarks only involve simple operations that are
easy to handle in \sparksql backend compiler and are expected to be fast.
Moreover, the parallel algorithms for these operations
fit reasonably well with the map-reduce
paradigm of \spark.
Nevertheless, \Sys is significantly faster since it uses HPC
technologies instead of relying on a distributed library.
Overall, the higher performance of \Sys enables interactive analytics for
 data scientists using a
small local cluster, since the execution times are in few seconds range
for queries with few operations.

The results for the filter benchmark provide insights about the trade-off between
generality and performance in these systems. Data frames of Python (Pandas) accept
any expression evaluating to Boolean array for filtering \dfs.
However, the expression is not
evaluated insides the optimized backend of Pandas and it can be slow.
On the other hand, \spark only allows simple expressions with hard-coded operations on the
data frame's columns (e.g. \CodeIn{df[`id']<100}). 
These expressions evaluate to a class of type ``Column'' which is used
for code generation in \sparksql backend.
\Sys provides best of both worlds using an end-to-end compiler approach.
It allows general array expressions as well as high performance execution simultaneously.

Figure~\ref{fig:basic} also demonstrates that parallel execution of aggregate in
\sparksql is slower than Python and Julia, while \Sys achieves significant (but relatively small) speedup.
 The reason is that the communication overheads are more dominant for this operation,
and using faster communication software stack is important.
Note also that these relational operations are data-intensive and one cannot expect to keep
all the processor cores busy, since memory bandwidth often becomes bottleneck.

\paragraph{Advanced Analytics Operations:}
We compare the performance of various systems for cumulative summation (\CodeIn{cumsum}),
simple moving average (SMA), and weighted moving average (WMA), which are common
analytics operations (Section~\ref{sec:syntax}).
 The input table has 256 million rows but the operation
is run on a single column.
Figure~\ref{fig:advanced} demonstrates that \Sys is 11$\times$, 837$\times$, and
15781$\times$ faster than Python (Pandas) for \CodeIn{cumsum}, SMA, and WMA, respectively.
In Python, SMA is significantly faster than WMA since SMA is
run in the optimized backend of Pandas whereas WMA requires the user to pass
a function that applies the weights. Again, this highlights the trade-off between
performance and flexibility in current systems,
while the end-to-end compiler approach of \Sys avoids this issue.

Furthermore, \Sys
is 1330$\times$, 15500$\times$, and 20356$\times$ faster than \sparksql.
These operations are fundamentally challenging for \sparksql since
they require communication operations other than the ones (e.g. reduction) that
are supported in map-reduce frameworks. 
Cumulative summation requires a scan (partial reductions) communication operation
while moving averages require near neighbor exchanges. Since \Sys is not
limited to \spark, it can generate the appropriate communication calls
(e.g. \CodeIn{MPI\_Exscan}). On the other hand, \sparksql gathers all the data
on a single executor (core) and performs the computation sequentially.
This results in data spill to disk even with the relatively modest dataset size used,
 due to excessive memory consumption of \spark.

\paragraph{User-defined functions (UDFs):}
The two language design of \sparksql has significant performance implications for many programs.
Data scientists often need custom operations which can be
written in the host language using \sparksql's UDF interface (similar to database systems).
These UDFs can slow down the program significantly because \sparksql only
compiles and inlines its own hardcoded operations (e.g. \CodeIn{+}, \CodeIn{*}, \CodeIn{exp}, ...).
On the other hand, the end-to-end compiler approach of \Sys naturally avoids this problem.

\begin{figure}
\begin{lstlisting}[language=Python,escapeinside={(*}{*)},mathescape,numbers=none]
# Spark SQL no-UDF version
df = spark.sql("SELECT id AS id, SUM(2*x) AS sx, SUM(2*y) AS sy FROM points GROUP BY id")

# Spark SQL UDF version
spark.udf.register("myudf",lambda x:2*x,DoubleType())
df = spark.sql("SELECT id AS id, SUM(myudf(x)) AS sx, SUM(myudf(y)) AS sy FROM points GROUP BY id")
\end{lstlisting}
\begin{lstlisting}[language=Julia,escapeinside={(*}{*)},mathescape,numbers=none]
# no-UDF version
df = aggregate(df, :id, :sx = sum(2*:x),
                        :sy = sum(2*:y))
# UDF version
myudf = x->2*x
df = aggregate(df, :id, :sx = sum(myudf(:x)),
                        :sy = sum(myudf(:y)))
\end{lstlisting}
\caption{\sparksql and \Sys versions of UDF performance benchmark.}
\label{fig:udf}
\end{figure}

We evaluate the performance impact of UDFs in \sparksql and \Sys using a simple benchmark we designed for
this purpose. For each system, we use a version with a UDF and a version without UDFs.
Figure~\ref{fig:udf} illustrates the performance impact of using different versions in \sparksql for Python and Scala interfaces and in \Sys.
The UDF version in \sparksql is 24\% slower with Python interface and 46\% slower with Scala interface.
However, the performance difference is negligible in \Sys since the generated codes are identical.

\begin{figure}
    \centering
    \def\svgwidth{0.8\columnwidth}
    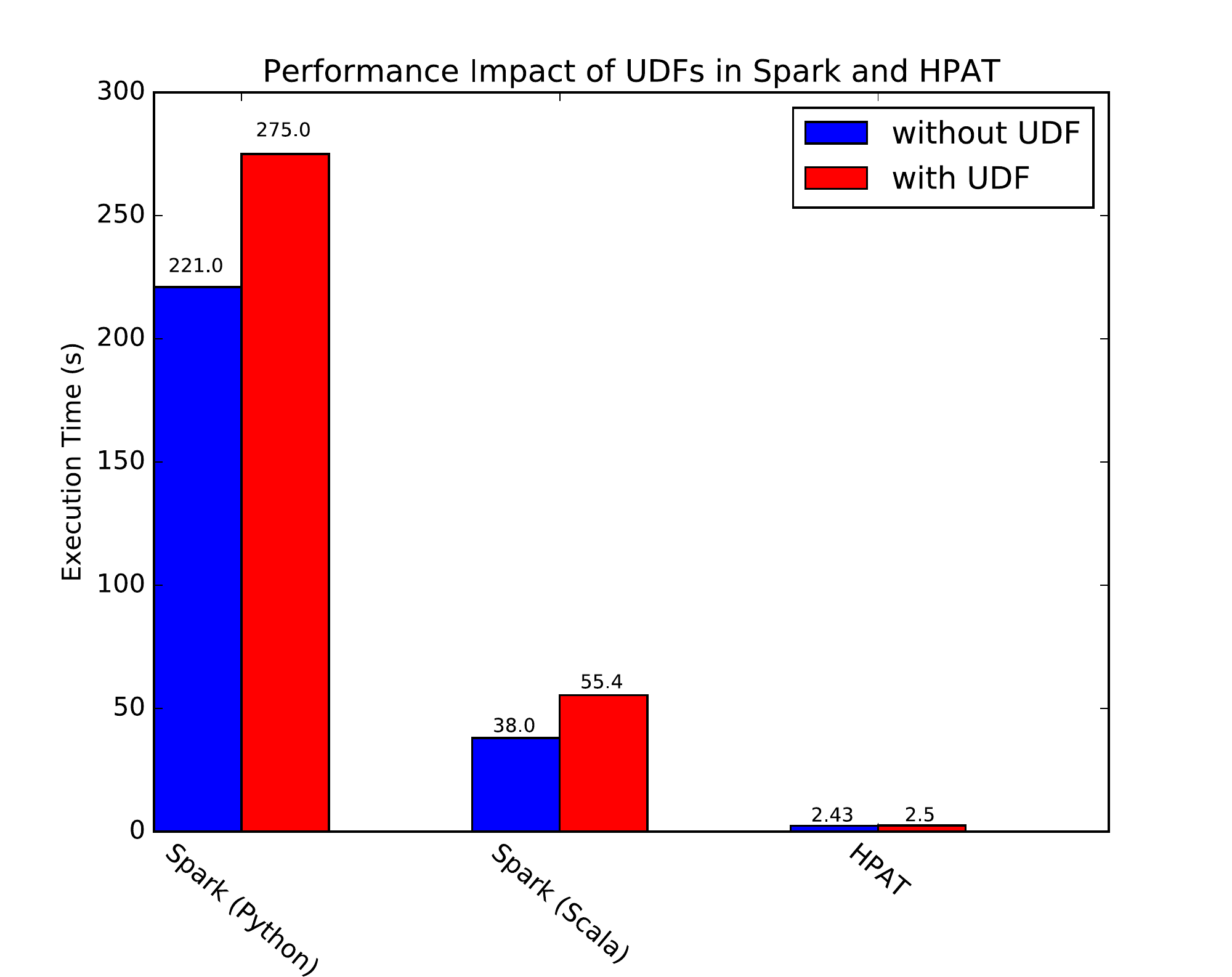
    \caption{Overhead of UDFs in \sparksql and \Sys.
    Using UDFs results in significant slowdown in \sparksql since there are two languages }
    \label{fig:udf}
\end{figure}

%
%
%

\begin{figure*}[t]
    \centering
\begin{subfigure}{.32\linewidth}
    \def\svgwidth{\columnwidth}
    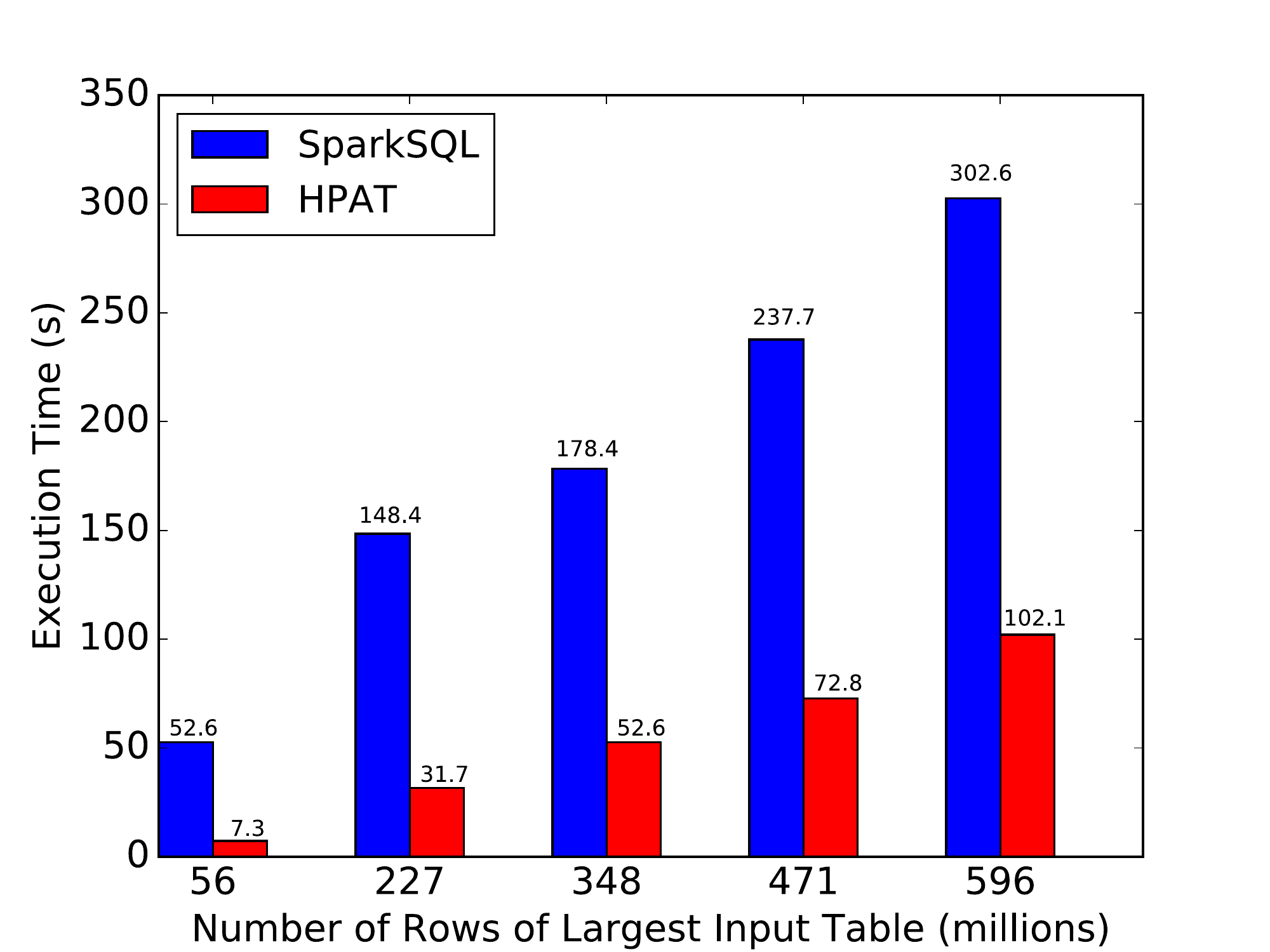
    \caption{Q26}
    \label{fig:q26}
\end{subfigure}
\begin{subfigure}{.32\linewidth}
    \def\svgwidth{\columnwidth}
    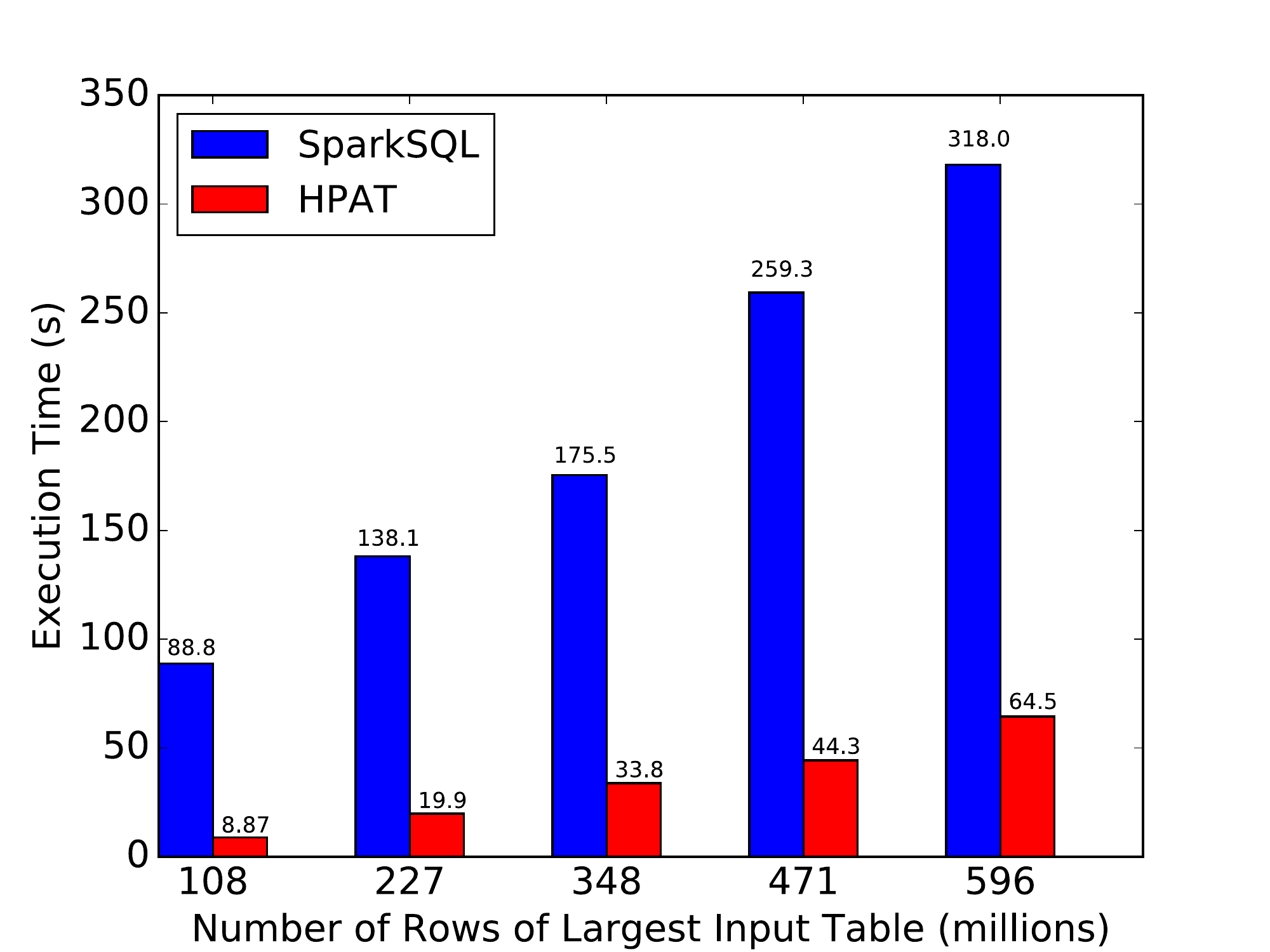
    \caption{Q25}
    \label{fig:q25}
\end{subfigure}
\begin{subfigure}{.32\linewidth}
    \def\svgwidth{\columnwidth}
    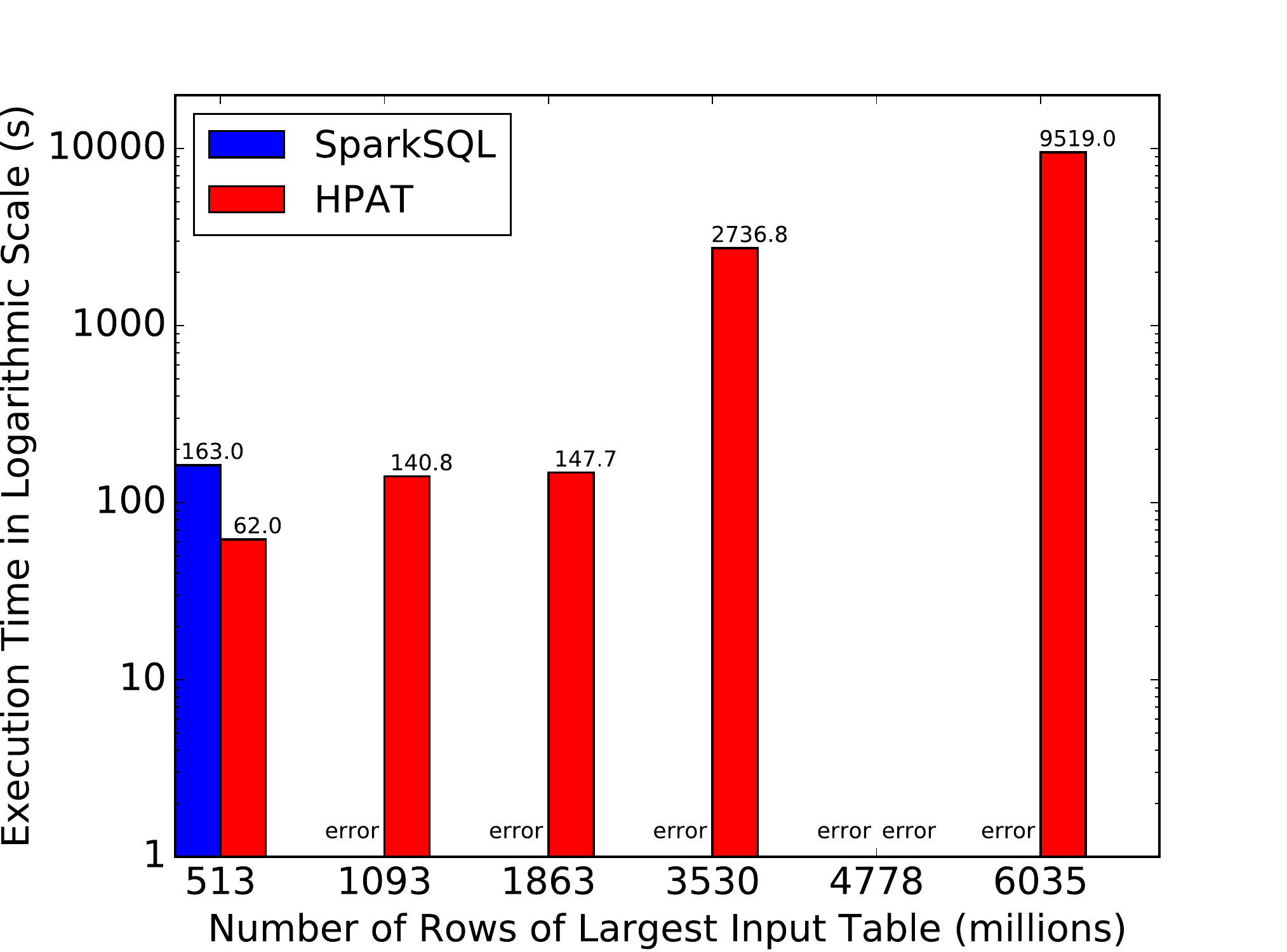
    \caption{Q05}
    \label{fig:q05}
\end{subfigure}
    \caption{Performance of TPCx-BB benchmarks in \sparksql and \Sys.}
\end{figure*}

\subsection{TPCx-BB Benchmarks}
To evaluate programs with multiple relational operations working together,
we use three benchmarks of TPCx-BB (BigBench)~\cite{Ghazal:2013}, which is designed
to evaluate the performance big data systems for analytics queries.
 We use Q05, Q25, and Q26 benchmarks since they
are tasks that include various stages from loading
 and transforming data to building training matrices and calling machine learning algorithms.
We use the default data generator in the suite~\cite{Rabl:2010:DGC}.
To focus on relational performance, we exclude data load time and machine learning algorithm execution times.
 The original benchmarks were written for Apache Hive
but we ported them to \sparksql
since \sparksql is shown to be significantly faster~\cite{armbrust2015}.
Even though we target scripting languages, we use the Scala/SQL
 interface for performance evaluation to observe the best performance of Spark.
 In addition, to enable uniform comparison across different problem sizes,
  we disable an optimization in
  Spark where tables smaller than a threshold are broadcast for join operations
  (\CodeIn{spark.sql.autoBroadcastJoinThreshold=-1}).
  This optimization is also not used in \Sys.
Note that since these benchmarks are designed for SQL systems, \sparksql is able to optimize
them easily, while they are stress tests for \Sys.

Figure~\ref{fig:q26} compares the performance of \sparksql and \Sys for Q26 benchmark of TPCx-BB,
demonstrating that
\Sys is 3$\times$ to 7$\times$ faster than \spark. One can conclude that \Sys
is capable of optimizing relational programs effectively in comparison to a SQL system.
Furthermore, \Sys is even faster since it employs a compiler and HPC
technologies rather than relying on a distributed library.


Figure~\ref{fig:q25} compares the performance of \sparksql and \Sys for Q25 benchmark of TPCx-BB and
demonstrates that \Sys is 5$\times$ to 10$\times$ faster than \sparksql.
The gap is wider for this benchmark partially because it requires more computationally expensive operations
(e.g. counting distinct values in aggregate) that benefit from low-level code
generation of \Sys.


Figure~\ref{fig:q05} compares the performance of \sparksql and \Sys for Q05 benchmark of TPCx-BB.
This benchmark is challenging since it involves a join on a large table with highly skewed
data. Hence, hash partitioning results in high load imbalance among processors which
is a well-known problem in the parallel database
literature~\cite{dewitt1992practical,xu2008handling,walton1991taxonomy}.
 \sparksql throws an error for scale factors greater than 50 since an internal sorting data structure
runs out of memory. \Sys throws an error for scale factor 400 only. The reason is a limitation
in current MPI implementations where the data item counts cannot be more than $2^{32}$.
Since load imbalance is well-studied in the HPC domain, \Sys could
take advantage of existing HPC technologies to address this problem. For example,
\Sys could generate code for \charm which includes advanced load balancing capabilities~\cite{sc14charm}
and naturally provides the {\it virtual processor} approach,
 which is proposed in the parallel database literature~\cite{dewitt1992practical}.
The implementation of this feature is left for future work.

\begin{figure}
    \centering
    \def\svgwidth{\columnwidth}
    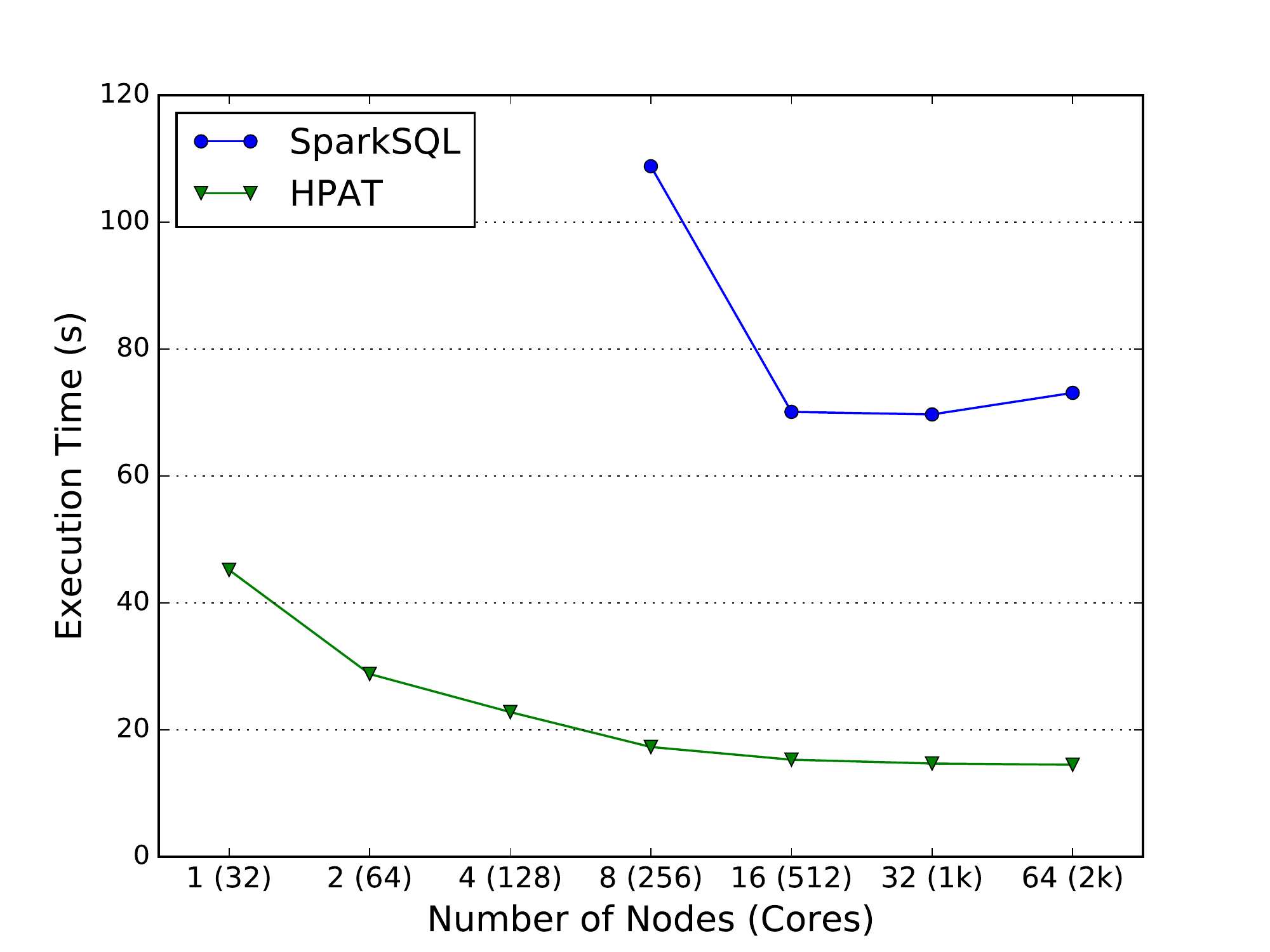
    \caption{Scaling of Q26 in \sparksql and \Sys.}
    \label{fig:scaling}
\end{figure}

\paragraph{Strong Scaling:} We compare scalability of \sparksql and \Sys for
large-scale distributed-memory machines using Cori (Phase I) supercomputer at NERSC~\cite{CORI}.
Each node has two Intel Xeon E5-2698 v3 processors (2$\times$16 cores) and
128GB of memory. Spark 2.0.0 is provided on Cori.
Figure~\ref{fig:scaling} demonstrates the strong scaling of
Q26 benchmark from one to 64 nodes (32 to 2048 cores).
We use scale factor 1000 dataset for this experiment,
where the larger input table has 1.2 billion rows.
Note that Spark crashes on settings with fewer than eight nodes because of
resource limitations.
The figure demonstrates that execution time of \Sys decreases using more nodes up to 64 nodes,
while \sparksql is slower on 64 nodes compared to 16 nodes.
Hence, \Sys is 5$\times$ faster than \sparksql on 64 nodes.
The fundamental scalability limitation of \spark is due to its master-slave
approach where the master becomes a sequential bottleneck~\cite{HPAT}.

\section{Related Work}\label{sec:related}
\Sys is the first compiler-based system for \dfs that automatically parallelizes
relational operations and tightly integrates with array computations.
Hence, this work is related to areas such automatic parallelization,
\sql embedding and big data processing.

\paragraph{Automatic Parallelization and Distribution:}
Automatic parallelization is extensively studied in the literature~\cite{Fonseca2016,7322481,6877466}, especially
in the context of High Performance Fortran (HPF)~\cite{Kennedy:2007:RFH:1238844.1238851,1437298}.
Typically, arrays in such systems are aligned to a {\it template} of
infinite virtual processors and then distributed based on heuristics.
The computation is then distributed based on the {\it owner-computes} rule.
For example, Kennedy and Kremer~\cite{Kennedy:1998:ADL:291891.291901} proposed a framework where
 the program is divided into phases (loop-nests)
 and all possible alignment-distribution pairs are found for each phase.
Then, performance models are used to evaluate various layout and remapping costs.
Finally, 0-1 integer programming is used to evaluate all possible layout and
remapping combinations for the whole program, which is an NP-complete problem.
However, auto-parallelization proved not to be practical because
the compiler analysis was too complex and the generated programs
significantly underperformed hand-written parallel programs~\cite{Kennedy:2007:RFH:1238844.1238851}.
\hpat solved the auto-parallelization problem
for scripting array programs in the data analytics and machine learning domain by exploiting domain knowledge~\cite{HPAT}.
We extend \hpat by integrating data frames and extending \hpat's parallelization
to relational operations. For example, we integrate a new parallelization
method in its semilattice of parallelism methods (Section~\ref{sec:dist-pass}).
Distributed Multiloop Language (DMLL) presents a new parallel IR and
 various transformations for heterogeneous platforms~\cite{Brown:2016:AEP:2854038.2854042}.
 However, DMLL starts from an explicitly parallel program and only includes simple
 parallelism inference for intermediate values.
 Nevertheless, some of their transformations could be used
 in \Sys for operations such as filter and aggregate, which is left for future work.


\paragraph{Distributed Library Approach:}
A common approach in previous work is building a \sql subsystem
on top of a distributed library, such as \sparksql~\cite{armbrust2015}
on top of \spark~\cite{zaharia2010} and DryadLINQ~\cite{dryadlinq} on top of Dryad~\cite{isard2007}.
These systems require writing relational code in
\sql or a \sql-like DSL, which is executed on top of the distributed library.
This approach has two main drawbacks for our target domain.
First, there is no tight integration with array computations in the rest of the
program. Second, the \sql subsystem inherits the fundamental performance limitations
 of these distributed libraries such master-slave
 and runtime scheduling overheads (see Section~\ref{sec:bigdatasys}).
\Sys avoids these issues by providing \df abstractions that are tightly integrated
with array computations, and are compiled to efficient parallel code.

\paragraph{Language Integrated Queries:}
The first and most straightforward method of accessing relational databases in a program was through the use of embedded \sql strings.
However, this approach has problems with type-safety, error checking and security~\cite{1553551,Maier:1990:RDP:101620.101642}.
Hence, language integrated queries have a long history ~\cite{Breazu-Tannen1992} and is still an area of active research~\cite{cheney2013practical,1553551,Maier:1990:RDP:101620.101642,linq2006}.
For example, Microsoft LINQ provides a DSL that is
equivalent to \sql and is integrated in .NET languages~\cite{linq2006}. Even though their
target domain is different, future work could potentially apply
some of LINQ's transformations in \Sys.
Language integrated queries are particularly popular in functional languages ~\cite{opaleye,Suzuki:2016:FSE:2847538.2847542,HRR,Cheney:2013:PTL:2500365.2500586} with their focus on type safety since integration allows queries to undergo compile-time type checking.  
Moreover, in some cases, these systems make it impossible by construction to create invalid SQL on the backend.
Finally, these systems tend not to support \dfs as such because their data stores are row-oriented rather than column oriented.



\section{Conclusion and Future Work}\label{sec:conclusion}
This paper introduced \Sys, which a compiler-based end-to-end
data analytics framework that integrates array computations and
relational operations seamlessly, and generates efficient parallel code.
We presented \Sys's \api, and the compiler techniques that make it possible.
Our evaluation demonstrated superior performance of \Sys compared to alternative systems.

\Sys opens various research directions. More compiler optimizations
across array computations and relational operations need to be explored.
In addition, generating faster parallel code and using HPC techniques such as
MPI/OpenMP hybrid parallelism could result in significant improvements.
Moreover, \Sys could potentially allow more complex data analytics programs
that need to be investigated.

\bibliographystyle{abbrvnat}
\bibliography{hpat}

\end{document}